\documentclass{IP_C}
\usepackage{url}

\IPcvolume{14}

\usepackage{algorithm,algorithmic}
\usepackage{graphicx}

\begin{document}

\title{Hiding inside Html and Other Source Codes}

%


\author{Hameed Al-Qaheri 
	\and  Sandipan Dey
	\and Sugata Sanyal}
\IPcAuthor{H. Al-Qaheri, S. Dey, S. Sanyal}
\institute{Department of Quantitative Methods and Information Systems,\\
		College of Business Administration \\
		Kuwait University \\
\texttt{alqaheri@cba.edu.kw} \\

Cognizant Technology Solutions\\
\texttt{sandipan.dey@gmail.com}\\ 

School of Technology and Computer Science,\\
Tata Institute of Fundamental Research,\\
Homi Bhabha Road, Mumbai - 400005, India\\
\texttt{sanyal@tifr.res.in}}

\date{}

\maketitle


\section*{Abstract}
\par Many steganographic techniques \cite{r1} \cite{r2} \cite{r3} \cite{r4} were proposed for hiding secret message inside images, 
the simplest of them being the LSB data hiding \cite{r6} \cite{r7} \cite{r8} \cite{r9} \cite{r10}, \cite{r11}. In this paper, we suggest a novel data hiding technique in an Html Web page \cite{r12} and also propose some simple techniques to extend the embedding technique 
to source codes written in any programming language (both case insensitive like html, pascal and case sensitive languages like C, C++, java) - an extension to \cite{r12}. We basically try to exploit the case-redundancy in case-insensitive language, while we try hiding data with minimal changes int the source code (almost not raising suspicion). Html Tags are case insensitive and hence an alphabet in lowercase and one in uppercase present inside an html tag are interpreted in the same manner by the browser, i.e., change in case in an web page is imperceptible to the browser. We first exploit this redundancy and use it to embed secret data inside an web page,
with no changes visible to the user of the web page, so that he can not even suspect about the data hiding.
The embedded data can be recovered by viewing the source of the html page. This technique can easily be extended to embed secret message inside any piece of source-code where the standard interpreter of that language is case-insensitive. For case-sensitive programming languages we do minimal changes in the source code (e.g., add an extra character in the token identified by the lexical analyser) without violating the lexical and syntactic notation for that language) and try to make the change almost imperceptible. 

\section{Introduction}
\par
Steganography is another name of hiding secret data in cover medium, thereby ensuring imperceptibility and exploiting redundancies in representation of the cover medium. For instance, in case of LSB data hiding the property that the cover image visual representation
is least affected (almost unaffected) by the change of the LSB of any pixel, is used and this redundancy (or cover image oblivious to change in LSB) is exploited to embed secret data in LSB \cite{r11}. Also, some decomposition techniques were proposed to enhance the LSB data hiding technique by increasing the number of bitplanes \cite{r6} \cite{r7} \cite{r8} \cite{r9}.
Some techniques for hiding data in executables are already proposed (e.g., Shin et al \cite{r4}). In this paper we introduce
a very simple technique to hide secret message bits inside source codes as well, as an extension of \cite{r12}. We describe our steganographic technique by hiding inside html source as cover text, but this can be extended to any case-insensitive language source codes like Basic, PASCAL or FORTRAN. 

\section{Hiding Data inside Html}

\begin{figure*} 
    \label{fig:f1}
    \centering
        \includegraphics[width=4.5cm, angle=270]{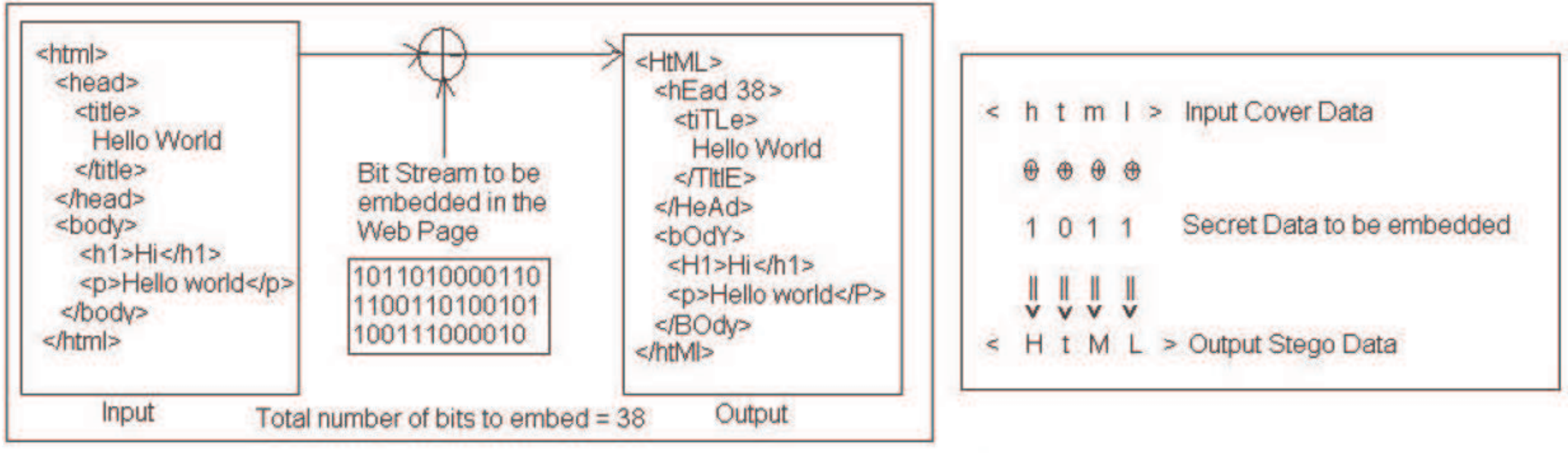}
    \caption{Illustration of how the Html data hiding works}
\hrule
\end{figure*}

\subsection{\hbox{Exploiting~the~Case-Insensitivity}}
As we know, Html Tags are basically directives to the browser and they carry information regarding how to structure and 
display the data on a web page. They are not case sensitive, so tags in either case (or mixed case) are interpreted
by the browser in the same manner (e.g., ``$<head>$'' and ``$<HEAD>$'' refers to the same thing). Hence, there is a redundancy 
in terms of case-insensitivity and we shall exploit this redundancy. To embed secret message bits into html, if the cases of the tag alphabets in html cover text are accordingly manipulated, then this tampering of the cover text will be ignored by the browser and hence it will be imperceptible to the user, since there will not be any visible difference in the web page, hence there will not be any suspect for it as well. Also, when the web page is displayed in the browser, only the text contents are displayed, not the tags (those can only
be seen when the user does `view source'). Hence, the secret messages will be kind of hidden to user, since they will have no effect 
on the page displayed by the browser. In other words, browser will help us hiding the data, by being indifferent to cases of the html tags, but we shall use those as key places for hiding data.

\par
\noindent Since only the portion of cover text inside the html tags will be used (and possibly will undergo a case-conversion) for hiding secret message bits and we are not going to tamper the html text data (outside the tags) that are going to be displayed by the browser as web page (this html cover text is analogical to the cover image, when thought in terms of steganographic techniques in images \cite{r1, r2, r3}), the user will not have any reason to suspect about hidden data in text. We shall only change the case of every character within these Html tags (elements) in accordance with the secret message bits that we want to hide inside the html source. \\
As described in \cite{r12}, if we think of the browser interpreter as a function, $f_B : \Sigma^{*} \rightarrow \Sigma^{*}$ we see that it is non-injective, i.e., not one to one, since $f_B(x)=f_B(y)$ whenever $x \in \{\mbox{`A'} \ldots \mbox{`Z'}\}$, $y \in \{\mbox{`a'} \ldots \mbox{`z'}\}$ and $Uppercase(y)=x$. The extraction process of the hidden message will also be very simple, one needs to just do `view source' and observe the case-patterns of the text within tags and can readily extract the secret message (and see the unseen), while the others will not know anything. So, both the embedding and extraction of secret message bits become very simple.

\par
\noindent As it can be guessed, we can not hide arbitrary long data inside a given cover text. More precisely, the length (in bits) of the secret message to be hidden inside the html cover-text will be upper-limited by the sum of size of text inside html tags (here we don't consider attribute values for data embedding. In case we consider attribute values for data embedding, we need to be more careful, since for some tags we should think of case-sensitivity, 
e.g. $<$A HREF=``link.html''$>$, since link file name may be case-sensitive on some systems, whereas, attributes such as
$<$h2 align=``center''$>$ is safe). If less numbers of bits to be embedded, we can embed the information inside Header Tag
specifying the length of embedded data (e.g. `$<$Header $25>$' if the length of secret data to be embedded is $25$ bits)
that will not be shown in the browser (optionally we can encrypt this integer value with some private key).
In order to guarantee robustness of this very simple algorithm one may use some simple encryption on the data to be embedded.

\subsection{The Algorithm for Hiding Data}
As described in \cite{r12}, the algorithm for embedding the secret message inside the html cover text is very simple and straight-forward. First, we need to separate out the characters from the cover text that will be candidates for embedding, these are the case-insensitive text characters inside Html tags.
Figure 2 shows a very simplified automata for this purpose.

\par \noindent Also, let us define the following functions before describing the algorithm:
\begin{itemize}
\item $l:\Sigma^{*} \rightarrow \Sigma^{*}$ as:
\par
$l(c) = \left\{
\begin{array}{c l}
  ToLower(c) & c \in \{\mbox{`A'}..\mbox{`Z'} \} \\
  c & otherwise
\end{array}
\right\}$
where $ToLower(c) = c + d$

\item $u:\Sigma^{*} \rightarrow \Sigma^{*}$ as:
\par
$u(c) = \left\{
\begin{array}{c l}
  ToUpper(c) & c \in \{\mbox{`a'}..\mbox{`z'}\} \\
  c & otherwise
\end{array}
\right\}$
where $ToUpper(c) = c - d$

\item Here $d = $ `A' $-$ `a'. \\
			The ascii value of `A' $=65$ and the ascii value of `a' $=97$, with  a difference $d = 32$.

\end{itemize}

\par \noindent It's easy to see that if the domain $\Sigma^{*}=\{\mbox{`a'}..\mbox{`z'}\} \cup \{\mbox{`A'}..\mbox{`Z'}\}$, then \\
$l:\{\mbox{`A'}..\mbox{`Z'}\} \rightarrow \{\mbox{`a'}..\mbox{`z'}\}$ and $u:\{\mbox{`a'}..\mbox{`z'}\} \rightarrow \{\mbox{`A'}..\mbox{`Z'}\}$,
implies that $l(.) = \overline{u(.)}=\Sigma^{*}-l(.)$. \\

\noindent Now, proceeding as in \cite{r12}, we want to embed secret data bits $b_{1}b_{2}..b_{k}$ inside the case-insensitive text inside the Html Tags. If $c_1 c_2 \ldots c_n$ denotes the sequence of characters inside the html tags in cover text (input html). A character $c_i$ is a candidate for hiding a secret message bit iff it is an alphabet. If we want to hide the $j^{th}$ secret message bit $b_j$ inside the cover text character $c_i$, the corresponding stego-text will be defined by the following function $f_{stego}$: 

$\forall{c_{i}} \in  \{\mbox{`a'}..\mbox{`z'}\} \cup \{\mbox{`A'}..\mbox{`Z'}\}$, i.e. if IsAlphabet($c_i$) is true,

$f_{stego}(c_{i}) = \left\{
\begin{array}{c l}
	l(c_{i}) & b_{j} = 0 \\
	u(c_{i}) & b_{j} = 1
\end{array}
\right\}$,

\noindent Hence, we have the following: 
\begin{eqnarray}
	\label{eq:}
    c_{i} \in  \{\mbox{`a'}..\mbox{`z'}\} \cup \{\mbox{`A'}..\mbox{`Z'}\} \Rightarrow f_{stego}(c_{i}) = l(c_{i}).\overline{b_{j}} \nonumber\\
+ u(c_{i}).b_{j}, \; \forall{i}
\end{eqnarray}

\par \noindent The number of bits ($k$) of the secret message embedded into the html cover text must also be embedded inside the html (e.g., in Header element). The figures 1, 2 and the algorithm \ref{alg:embed} together explain this data hiding algorithm.
\begin{algorithm} [h]
\begin{algorithmic}[1]
\STATE Search for all the html tags present in the html cover text and extract all the characters $c_1 c_2 \ldots c_n$ from
inside those tags using the DFA described in the figure 2.
\STATE Embed the secret message length $k$ inside html header in the stego text.
\STATE $j \leftarrow 0$.
\FOR {$c_i \in HTMLTAGS, \; i=1 \ldots n$}
\IF {$c_i \in \{\mbox{`a' \dots `z'}\} \cup \{\mbox{`A' \dots `Z'}\}$}
\STATE $f_{stego}(c_{i}) = l(c_{i}).\overline{b_{j}} + u(c_{i}).b_{j}$.
\STATE $j  \leftarrow j + 1$.
\ELSE
\STATE $f_{stego}(c_{i}) = c_{i}$.
\ENDIF
\IF {$j == k$}
\STATE break.
\ENDIF
\ENDFOR
\end{algorithmic}
\caption{Algorithm to Hide Data inside Html} \label{alg:embed} \vspace{.06 in}
\end{algorithm}
\begin{figure*} 
    \label{fig:f2}
    \centering
        \includegraphics[width=9cm,angle=270]{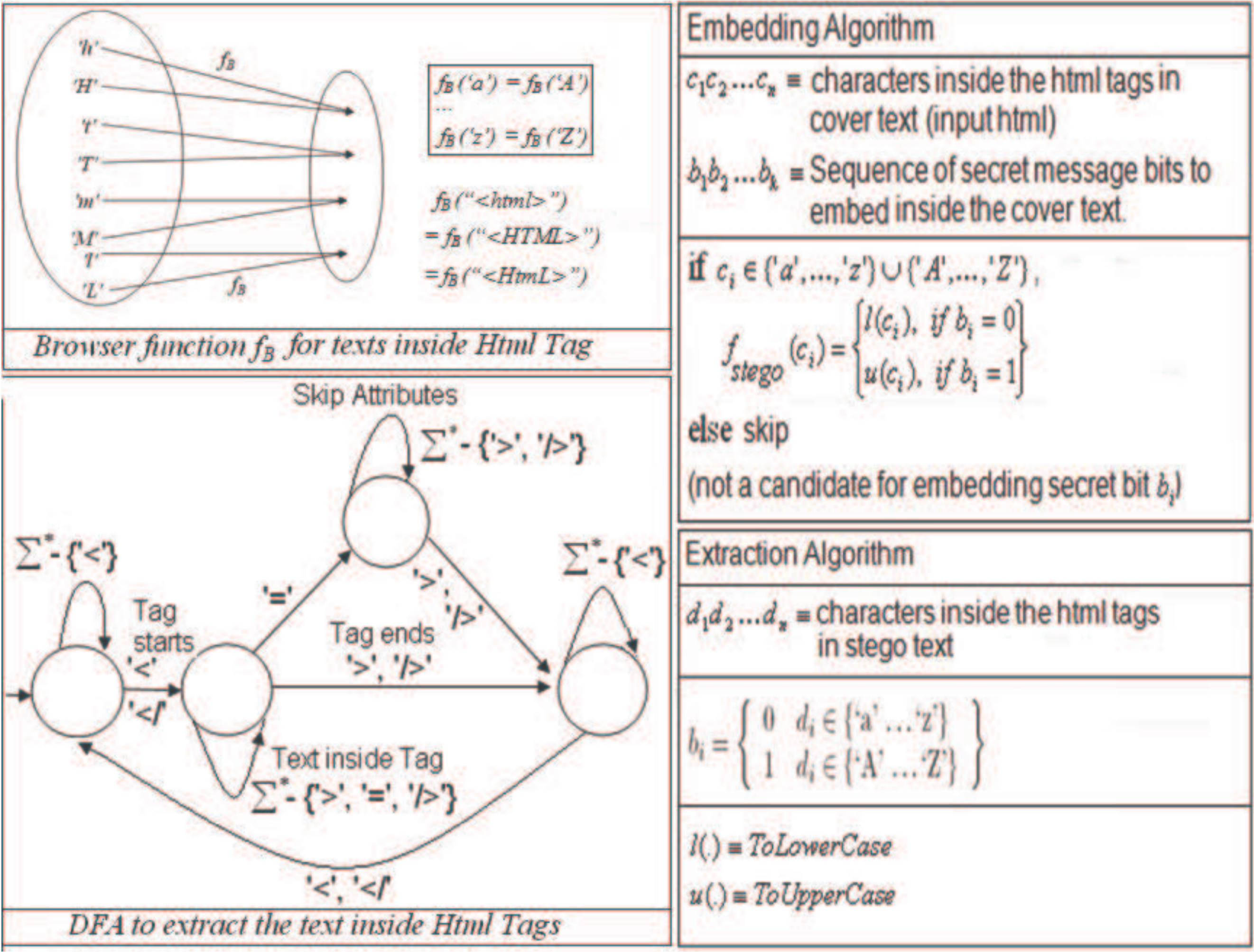}
    \caption{Basic block-diagram for the Html data-hiding technique}
\hrule
\end{figure*}

\subsection{The Algorithm for Hidden Data Extraction}
Again, proceeding as in \cite{r12}, the algorithm for extraction of the secret message bits will be even simpler. As in the embedding process, we must first separate out the candidate text (exactly the text within the Html tags) that were chosen in the earlier step for embedding secret message bits. Also, we must extract the number of bits ($k$) embedded into this page (e.g., from the header element). In order to find out the stego-text, one has to use `view source'.

\begin{algorithm} [h]
\begin{algorithmic}[1]
\STATE Search for all the html tags present in the html stego text and extract all the characters $d_1 d_2 \ldots d_n$ from
inside those tags using the DFA described in the figure 2.
\STATE Extract the secret message length $k$ from inside html header in the stego text.
\STATE $j \leftarrow 0$.
\FOR {$d_i \in HTMLTAGS, \; i=1 \ldots n$}
\IF {$d_i \in \{\mbox{`a' \dots `z'}\}$}
\STATE $b_{j} = 0$.
\STATE $j  \leftarrow j + 1$.
\ELSIF {$d_i \in \{\mbox{`A' \dots `Z'}\}$}
\STATE $b_{j} = 1$.
\STATE $j  \leftarrow j + 1$.
\ENDIF
\IF {$j == k$}
\STATE break.
\ENDIF
\ENDFOR
\end{algorithmic}
\caption{Hidden Data Extraction Algorithm} \label{alg:extract} \vspace{.06 in}
\end{algorithm}

\par
\noindent Now, we have $d_{i} = f_{stego}(c_{i}), \; \forall{i} \in \{1,2,\ldots,n\}$. If $d_i \in \{\mbox{`a'}..\mbox{`z'}\} \cup \{\mbox{`A'}..\mbox{`Z'}\}$ i.e., an alphabet, then only it is a candidate for decoding and to extract $b_{i}$ from $d_{i}$, we use the following logic: \\
\par\noindent
$b_{i} = \left\{
\begin{array}{c l}
	0 & d_{i} \in \{\mbox{`a' \dots `z'}\} \\
	1 & d_{i} \in \{\mbox{`A' \dots `Z'}\}
\end{array}
\right\}$ \\
\par \noindent Repeat the above algorithm $\forall{i} < k$, to extract all the hidden bits.
\subsection{Experimental Results}
As in \cite{r12}, we obtained the following results:\\
\par \noindent The figures 3, 4 and 5 (as in \cite{r12}) show an example of how our method works, while Figure 6 shows the comparison of
the histogram of the cover html and stego html in terms of the (ascii) character frequencies. Classical image hiding techniques
like LSB data hiding technique always introduce some (visible) distortion \cite{r5, r10} in the stego image (that can be reduced using techniques \cite{r6, r7, r8, r9}), but our data hiding technique in html is novel in the sense that it introduces no visible distortion in stego text at all.
\begin{figure*} [p]
    \label{fig:htmlsrc-cover}
    \centering
        \includegraphics[width=14cm,height=18cm]{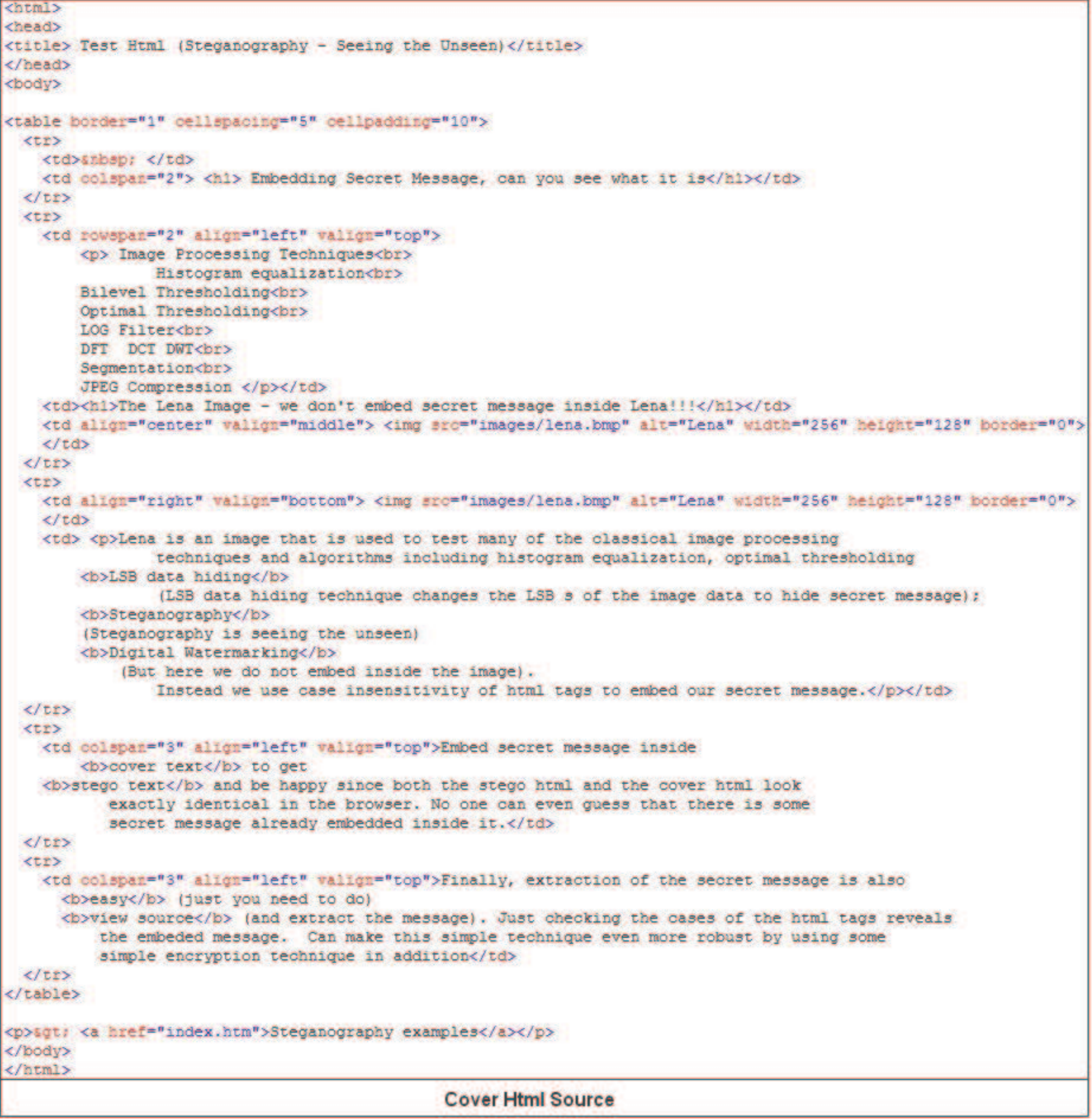}
    \caption{Cover Html source before embedding the secret message}
\end{figure*}

\begin{figure*} [p]
    \label{fig:htmlsrc-stego}
    \centering
        \includegraphics[width=14cm,height=18cm]{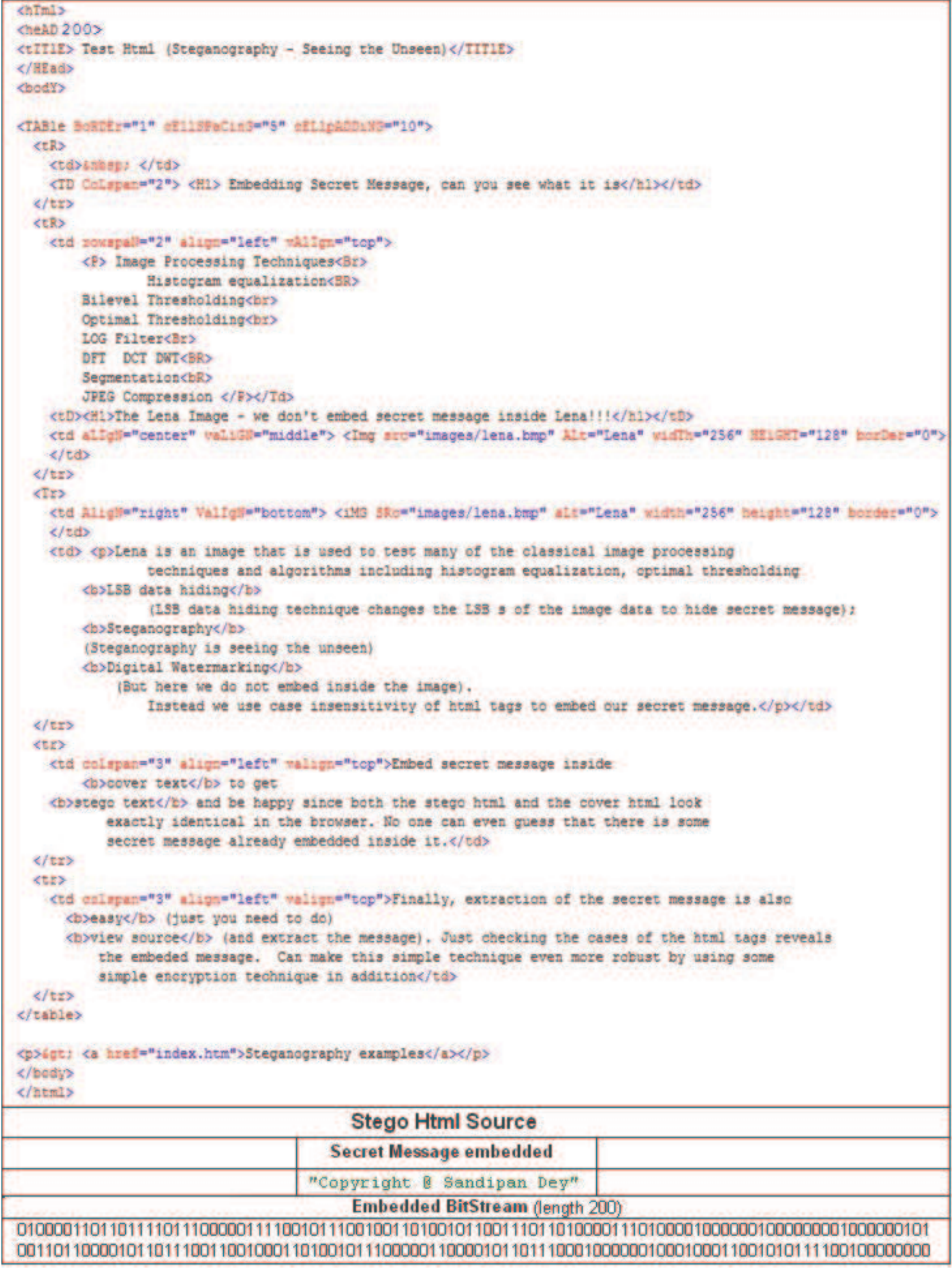}
    \caption{Stego Html source after embedding the secret message}
\end{figure*}

\begin{figure*} 
    \label{fig:html}
    \centering
        \includegraphics[width=12cm, angle=270]{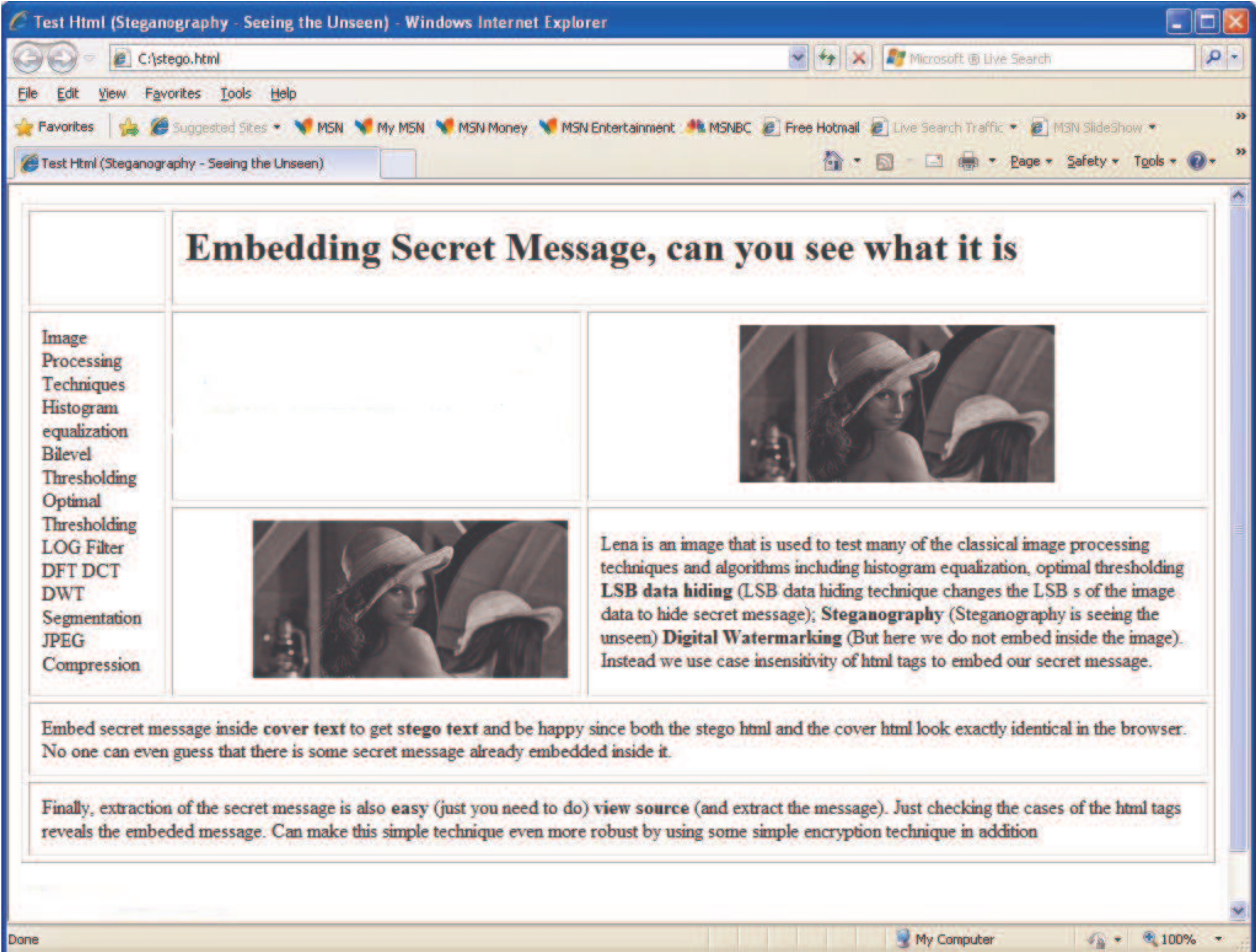}
    \caption{Cover $\&$ Stego Html}
\end{figure*}

\begin{figure*} 
    \label{fig:graph}
    \centering
        \includegraphics[width=12cm]{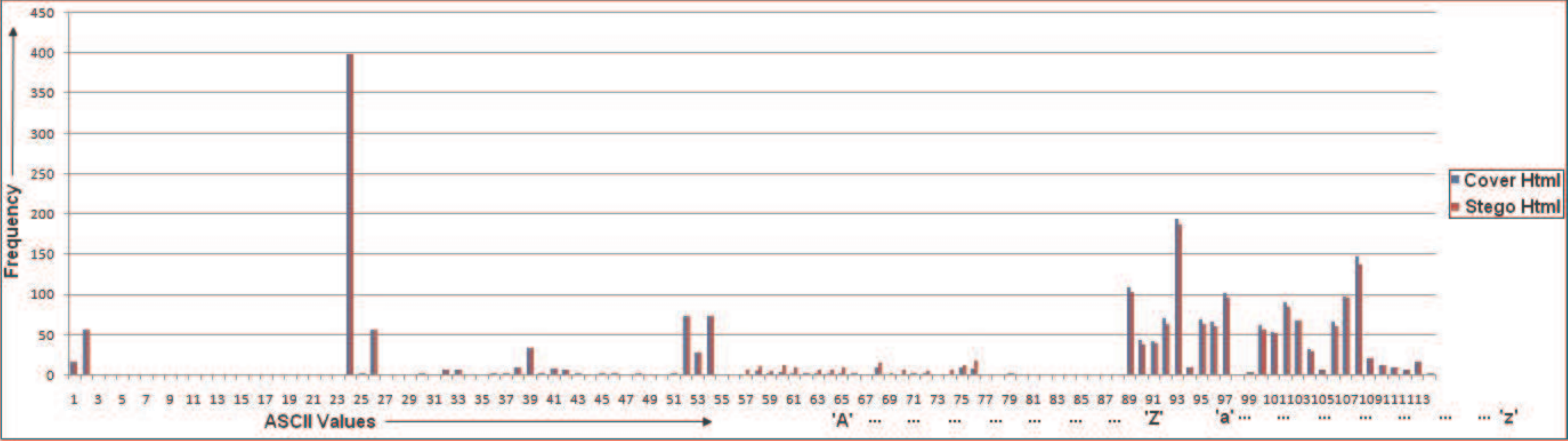}
    \caption{Cover vs Stego Html Histogram}
\end{figure*}
\begin{figure*}[t]
    \label{fig:dfa}
    \centering
        \includegraphics[width=6cm, angle=270]{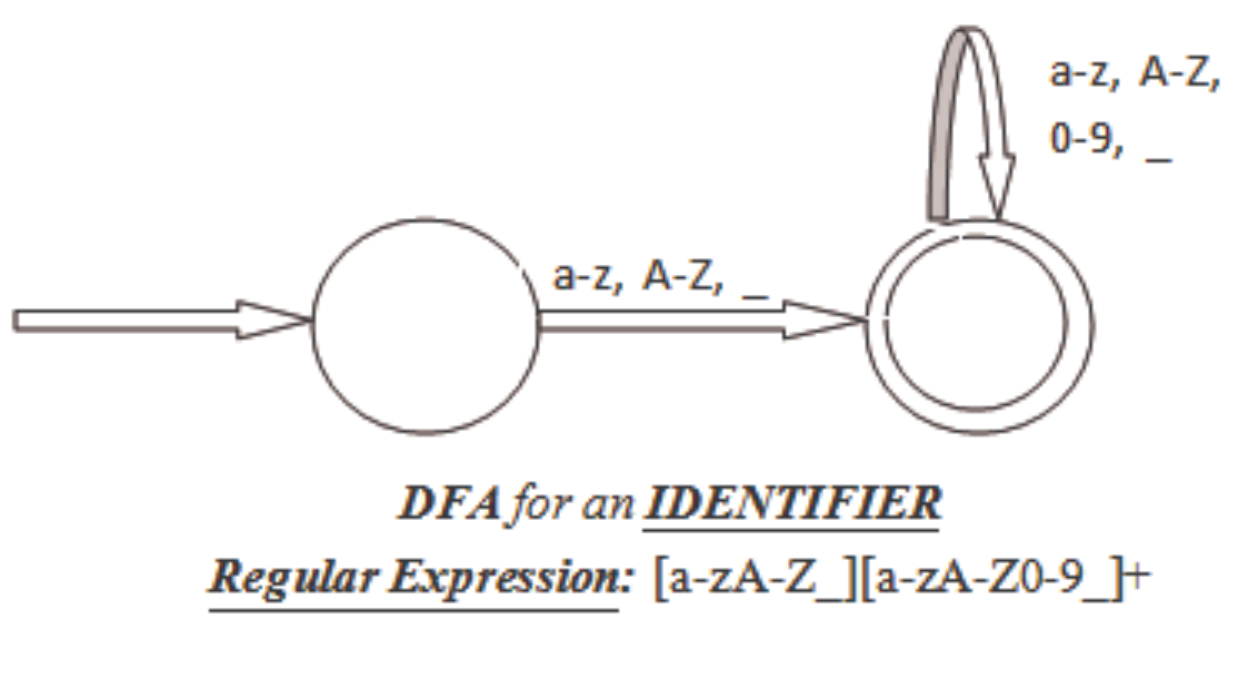}
    \caption{DFA for accepting IDENTIFIER}
\hrule
\end{figure*}

\section{Hiding Data inside Other Source Codes}
\subsection{Case-insensitive Programming Language Sources}
In order to hide data in the source codes in languages for which the letter-case is ignored (e.g. PASCAL), we can follow similar approaches as before. But we must ensure the following things:
\begin{itemize}
\item While embedding secret messages we must not change anything in the source code, so that the output of the program (when run)
changes.
\item Should not embed inside variable values, for instance string literals.
\item Should store the total number of bits embedded inside the source code somewhere, with or without encryption. 
\end{itemize}
There can be a few very simple ways of embedding and depending upon where to embed secret data bits there can be a few variants accordingly:
\begin{itemize}
\item Embed aggressively at every possible places (except possibly inside variable values, constants or string literals), in every keyword and identifier.
\item do not use all the characters of a candidate word for embedding, instead only use the first character (change the case depending
upon the next secret bit to be embedded, keeping all other characters unchanged).
\item Only embed inside the keywords.
\item Only embed inside the identifiers.
\end{itemize} 

\subsection{Case-sensitive Programming Language Sources}
Languages like C are case-sensitive, we can not use the above mentioned techniques directly. Off course we can hide data inside comments,
but what if a source does not have a comment at all? Creating some arbitrary artificial comments and embedding data inside them may not be a good idea. Instead we can use the following simple general technique: 
\begin{itemize}
\item Use lexical analyzer (some simple scanner for the language) to find IDENTIFIER tokens ($IDENTIFIER: [a-zA-Z\_][a-zA-Z0-9\_]+$, as shown in figure 7).
\item Only use variable names to embed secret data, no function name. Also, stick to local/static variables and do not use extern variable names to embed), use some simple parser to achieve this.
\item If next message bit to embed is 1, change the identifier name to append (or prepend) by an underscore($\_$), otherwise leave it as it is (e.g., if the variable name is \textit{var}, change it to \textit{var$\_$}, if the next bit to embed is 1, otherwise keep it as it is, while extraction interpret in the same manner).
\item Use some kind of symbol table (hash map) to keep track of every change in identifier name and accordingly reflect the change to all places where the identifier is used.
\item Skip compiler directives / Macros/ keywords.
\item Keep track of total number of bits embedded (e.g., store it inside a beginning comment, with / without simple encryption).
\end{itemize}
%
\section{Conclusions}
In this paper we presented simple algorithms and techniques for hiding data in html text and other source codes. 
This technique can be extended to any case-insensitive language and data can be embedded in the similar manner, 
e.g., we can embed secret message bits even in source codes written in languages like basic or pascal or in the 
case-insensitive sections (e.g. comments) in C like case-sensitive languages. Even for C-like case sensitive 
languages we can embed with minimal distortion by tweaking for instance the identifier name a little bit. 
Data hiding methods in images results distorted stego-images, but data hiding technique proposed for html 
does not create any sort of visible distortion in the stego html text.

\end{document}